\documentclass[11pt, a4paper]{article}
\usepackage[utf8]{inputenc}
\usepackage[left=2cm, right=2cm, bottom=2cm, top=2cm]{geometry}
\usepackage{color}
\usepackage{xcolor}
\usepackage{bm}
\usepackage{amsmath, amsfonts, amssymb}
\usepackage{siunitx}
\DeclareSIUnit{\angstrom}{\textup{\AA}}
\usepackage{url}
\usepackage{svg}
\usepackage{booktabs}
\usepackage{graphicx}
\usepackage{caption}
\definecolor{amethyst}{rgb}{0.6, 0.4, 0.8}
\definecolor{revcol}{rgb}{0,0,1}
\usepackage[colorlinks, citecolor=blue, linkcolor=teal, urlcolor=amethyst]{hyperref}
\usepackage[capitalise, nameinlink]{cleveref}
\usepackage{authblk}
\usepackage[square, numbers, comma, sort&compress]{natbib}

\renewcommand{\vec}[1]{\mathbf{#1}}

\usepackage{dcolumn}% Align table columns on decimal point
\usepackage{mathtools}
\usepackage{float}

\title{A universal attractive interaction between filaments at the cell scale}
\author{Anne-Florence Bitbol\textsuperscript{1,2,*}, Hélène Berthoumieux\textsuperscript{3,4}, Benjamin Spreng\textsuperscript{5}, Paulo A. Maia Neto\textsuperscript{6,*},  Serge Reynaud\textsuperscript{7}}
\affil{\textbf{1} Institute of Bioengineering, School of Life Sciences, École Polytechnique Fédérale de Lausanne (EPFL), CH-1015 Lausanne, Switzerland\\
\textbf{2} SIB Swiss Institute of Bioinformatics, CH-1015 Lausanne, Switzerland\\
\textbf{3} {Sorbonne Universit\'e, CNRS, Laboratoire de Physique Th\'eorique de la Mati\`ere Condens\'ee, Campus Jussieu, 
F-75005 Paris, France}\\
\textbf{4} {Fachbereich Physik, Freie Universität Berlin, Arnimallee 14, Berlin, 14195, Germany}\\
\textbf{5} Department of Electrical and Computer Engineering, University of California, Davis, CA 95616, USA\\
\textbf{6} Instituto de F\'{\i}sica, Universidade Federal do Rio de Janeiro \\ Caixa Postal 68528,   Rio de Janeiro,  RJ, 21941-972, Brazil\\
\textbf{7} Laboratoire Kastler Brossel, Sorbonne Universit\'e, CNRS, ENS-PSL, Coll\`ege de France, Campus Jussieu, F-75005 Paris, France \\
* Corresponding authors: \href{mailto:anne-florence.bitbol@epfl.ch}{anne-florence.bitbol@epfl.ch}, \href{mailto:pamn@if.ufrj.br}{pamn@if.ufrj.br}}
\date{\today}

\begin{document}

\maketitle

\begin{abstract}
Actin filaments and microtubules both often form bundles of parallel filaments within cells. Here, we shed light on a universal attractive interaction between two such parallel filaments. Indeed, the electrodynamic Casimir interaction between dielectric objects immersed in salted water at room or body temperature includes a universal contribution that is unscreened by the solvent and therefore long-ranged. We study this interaction between two parallel cylinders immersed in salted water with strong Debye screening. We show that its magnitude can largely exceed the energy scale of thermal fluctuations in the case of actin filaments and microtubules in cells. While multiple interactions exist between filaments in cells, this universal attractive interaction should thus have an important role, e.g. in bundle formation and cohesion. 
\end{abstract}

\section*{Introduction}

Filamentous structures are ubiquitous in cells. Cytoskeletal filaments, in particular actin filaments and microtubules in eukaryotes, play crucial parts in maintaining the integrity of cell shape, in its deformations, as well as in multiple sub-cellular processes. Actin filaments and microtubules form dynamic networks in cells, permanently undergoing polymerization and depolymerization in active processes requiring ATP hydrolysis. They are key components of the cytoskeleton, which mechanically supports eukaryotic cells, and actively generate forces with the help of motor proteins~\cite{Salbreux12,Murrell15,Burla19}. Actin filaments form bundles, where filaments are cross-linked by specific proteins into parallel arrays. Parallel bundles support projections of the cell membrane such as microvilli, microspikes or filopodia, while contractile bundles are present in stress fibers, and in the contractile ring that divides cells in two at mitosis. Microtubules, which are thicker and more rigid than actin filaments, also form bundles of parallel microtubules cross-linked by microtubule-associated proteins~\cite{Balabanian18}. Plant cells often possess large arrays of parallel microtubules, whose alignment is maintained over the whole cell. Microtubule bundles are also involved in the mitotic spindle of yeast and mammals~\cite{Gaillard08}. Long microtubule bundles are prominent in neurons' axons and dendrites too~\cite{Chen92}. Both in the case of actin filaments~\cite{Tang96,Deshpande12} and in that of microtubules~\cite{Needleman04,Hamon11,Chung16}, bundles have been shown to form \textit{in vitro} in the absence of cross-linkers under certain experimental conditions. Beyond the cytoskeleton, several enzymes form filaments in cells, with important biological functions, and these filaments also often self-assemble into larger assemblies, especially bundles~\cite{Park19}. 

Here, we study the electrodynamic Casimir interaction between two parallel filaments immersed in salted water with strong Debye screening, typical of the cytoplasm. We show that it constitutes a quantitatively important universal interaction at the cell scale. It was recently found that the Casimir interaction between dielectric objects immersed in salted water includes an attractive long-range contribution that is unscreened by the surrounding solvent \cite{MaiaNeto2019,Pires2021,Schoger2022}. This long-range Casimir interaction is particularly important in the geometry of two parallel cylinders. Indeed, the associated binding energy is proportional to the length of the cylinders, which can be much larger than their radii. This geometry of long cylinders is relevant for filaments at the cell scale. We show in the present letter that this attractive interaction substantially exceeds the thermal energy for actin filaments and for microtubules at the physiological separations found in bundles. We discuss implications for actin filaments, microtubules and other filaments.

\section*{Results}

\label{sec:res}

We study the electrodynamic Casimir interaction at room  or body temperature between two filaments immersed in an aqueous solution modeling the cytoplasm. The ion content of the cytoplasm entails that electrostatic interactions between them are screened, down to a separation of the order of $0.5\,{\rm nm}$~\cite{Spitzer05}. Despite this strong screening, the Casimir interaction includes a long-range non-screened part, due to the effect of thermal electrodynamical fluctuations propagating in the medium without being screened~\cite{MaiaNeto2019}. To illustrate this key point, while relying on a molecular description of the environment, we perform molecular dynamics simulations, supported by a classical field theory calculation (see Supplementary Material). We simulate pure water as well as an electrolyte solution with concentration 0.2 mole per liter, in the range of typical cytoplasmic concentrations~\cite{kornyshevtrans}.  We compute the dielectric correlation spectrum in these two media. We thereby confirm that the longitudinal spectrum is modified by the presence of salt, due to the screening of the correlations in electrolytes beyond the Debye length, whereas the transverse correlation spectrum remains unaffected by the presence of salt. The absence of screening of the transverse modes by salt allows the long-range Casimir force.

\begin{figure}[htbp]
    \centering
    \includegraphics[width=0.5\textwidth]{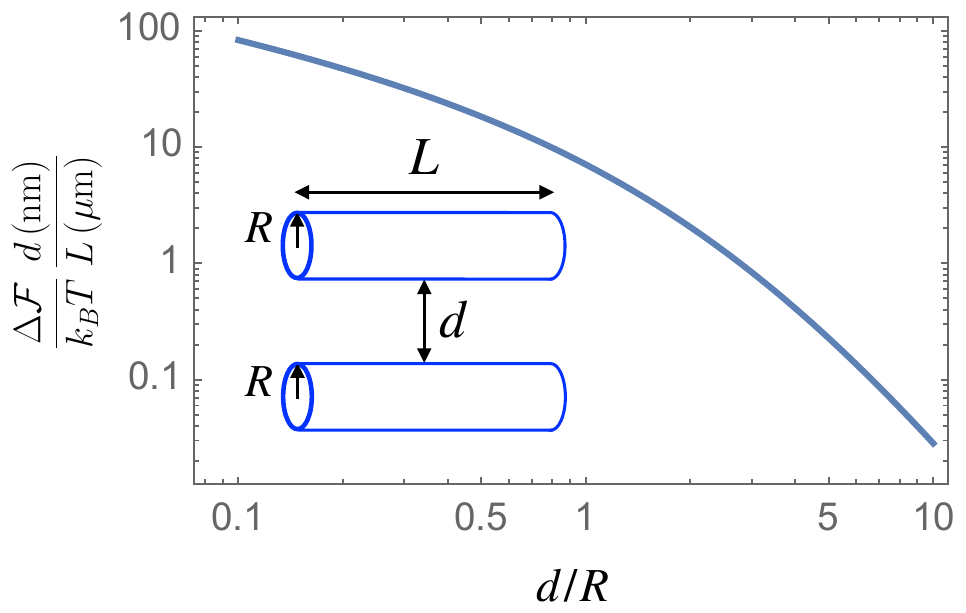}
    \caption{
    \textbf{A universal attractive interaction between two parallel filaments. } 
    The Casimir binding free energy measured in units of $k_BT$, namely $\Delta\mathcal{F}/(k_BT)$, is obtained by multiplying the ratio $L/d$ of the cylinder length $L$ to the separation distance $d$ by the universal function $\phi$, see Eq.~\ref{eq:deff}. With typical values of $L$ and $d$ measured respectively in micrometers and nanometers, the magnitude of the Casimir binding free energy is given by the quantity $\Delta\mathcal{F}/(k_BT)\times d(\mathrm{nm})/L(\mu\mathrm{m})=10^3\phi$. This quantity is plotted on the figure versus the ratio $d/R$ of the separation distance $d$ to the cylinder radius $R$. Inset: sketch of the geometry considered.}
    \label{fig:fig1}
\end{figure}

The universal long-range Casimir interaction in salted water was recently calculated between planes~\cite{MaiaNeto2019} and between spheres \cite{Schoger2022}. Furthermore, an experiment between spheres immersed  in salted water \cite{Pires2021} confirmed the theoretical predictions.
Here, we focus on the geometry of two long parallel filaments, modeled as dielectric cylinders with radius $R$ and length $L$, separated by a distance $d$ (see Methods and \cref{fig:fig1}). The Casimir interaction is expected to play a larger role than between two spheres, because it is proportional to the cylinder length $L$, which is much longer that their radius $R$. We summarize the calculations, based on the scattering theory of the Casimir interaction \cite{Lambrecht2006,Rahi2009}, in the Methods. The key result is that there exists a universal function $\phi$, which only depends on the aspect ratio $d/R$ characterizing the radial geometry, such that the Casimir binding free energy $\Delta\mathcal{F}$ between two filaments reads
\begin{eqnarray}\label{eq:deff}
\Delta\mathcal{F}= k_B T\,\frac{L}{d}\,\phi\!\left(\frac{d}{R}\right)~,
\end{eqnarray}
where $k_B T$ is the energy scale of thermal fluctuations, with $k_B$ denoting the Boltzmann constant and $T$ the absolute temperature. For simplicity, we take the convention that a positive binding free energy $\Delta\mathcal{F}$ means an attractive interaction, with $\Delta\mathcal{F}$ representing the variation of free energy upon unbinding. Remarkably, apart from the thermal energy, which sets the relevant scale for the electrodynamic Casimir free energy at room or body temperature~\cite{Schoger2022}, only the geometry of the system is involved in Eq.~\ref{eq:deff}, via the distances $L$, $d$ and $R$. In particular, the material properties of the filaments or of the surrounding cytoplasm do not enter it. While the mathematical expression of $\phi$ is complex, we compute it using a numerical method extending to the geometry of two cylinders what has already been done for planes or spheres (see Methods and Code Availability). For actin filament bundles and for microtubule bundles, the filament length $L$ will be in the micrometer range while the inter-filament distance $d$ will be in the nanometer range. 

In \cref{fig:fig1}, we plot $\Delta\mathcal{F}/(k_B T)\times d/L$, with $L$ and $d$ measured in~$\mu$m and in~nm, respectively, versus the ratio $d/R$. It shows that, for the values of $L$ and $d$ typical of the biological objects discussed below (see Figs.~2-3), $\Delta\mathcal{F}$ can take values substantially larger than $k_B T$. Therefore, the non-screened contribution to the Casimir force is highly relevant in the case of filaments at the cell scale.

\begin{figure}[htbp]
    \centering
    \includegraphics[width=0.5\textwidth]{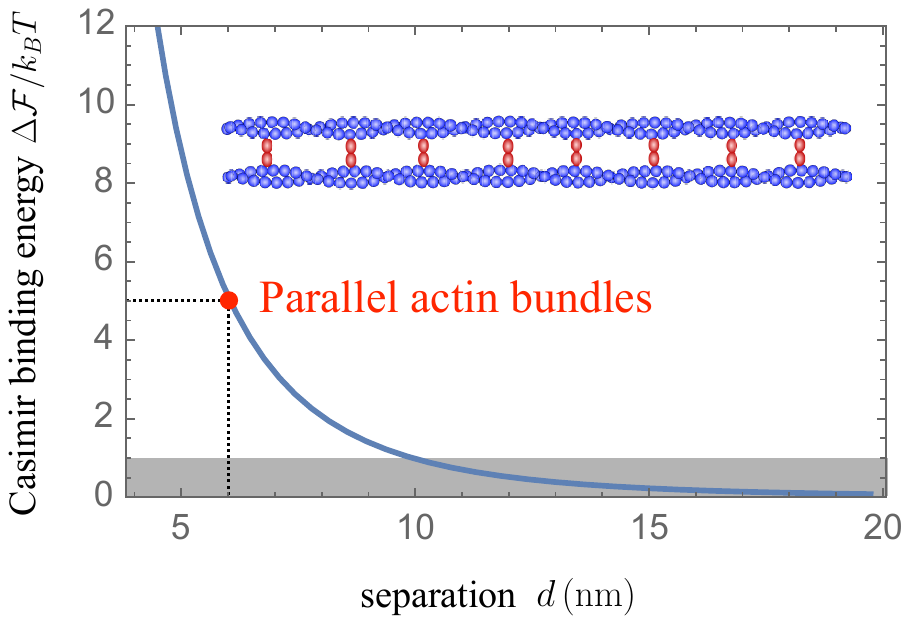}
    \caption{ \textbf{Casimir binding free energy between actin filaments.}
   The Casimir binding free energy $\Delta\mathcal{F}$ (from Eq.~\ref{eq:deff}) is shown versus the separation distance $d$~(nm) for two parallel actin filaments. Each filament has a radius $3\,{\rm nm}$ and length $15\,\mu{\rm m}$. $\Delta\mathcal{F}$ is expressed in units of $k_BT$, and the zone where $\Delta\mathcal{F}<k_BT$ is shaded in gray. Casimir binding free energies above the gray zone are expected to matter in practice, including in the physiological case of parallel actin bundles (red marker). Inset: schematic of two actin filaments (blue) in a parallel actin bundle. Cross-linkers are shown in red.
}
    \label{fig:fig2}
\end{figure}

Next, we assess more precisely the magnitude of this interaction in the concrete case of actin bundles. Actin filaments are double helices of homopolymers of monomeric actin. They can be approximately described as cylinders with a radius $R$ around 3~nm. They form bundles in cells, where actin filaments are cross-linked by specific proteins into arrays of parallel filaments. In parallel actin bundles, which support projections of the cell membrane such as microvilli, microspikes or filopodia, actin filaments (assembled with fimbrin, fascin or villin) are approximately $d=6$~nm apart (note that we use the closest approach distance $d$ here and throughout)~\cite{Claessens08,Volkmann01}. In this case, Eq.~\ref{eq:deff} yields a Casimir binding free energy per unit length of $\Delta{\cal F}/L = 0.33\, k_B T /\mu{\rm m}$, giving the substantial value $\Delta{\cal F}=5\, k_B T$ for a length $L= 15\,\mu{\rm m}$, which is on the order of the size of a cell and of the persistence length of actin filaments~\cite{Brangwynne07}. Such a value, significantly larger than the scale $k_B T$ of thermal fluctuations, demonstrates the practical relevance of the Casimir interaction between actin filaments in the physiological configuration of parallel bundles. In contractile bundles, which are present in stress fibers, and in the mitotic contractile ring, actin filaments (assembled with alpha-actinin) are separated by $d=33$~nm~\cite{Volkmann01}, yielding a smaller value of $\Delta{\cal F}\sim 10^{-2}\, k_B T$ for a length $L= 15\,\mu{\rm m}$, which is not relevant as it is well below the scale $k_B T$ of thermal fluctuations. In \cref{fig:fig2}, we show the Casimir binding free energy $\Delta\mathcal{F}$ versus the separation $d$ for actin filaments with $L= 15\,\mu{\rm m}$. The Casimir interaction then exceeds the scale of thermal fluctuations for separations $d\lesssim 10$~nm, which includes parallel bundles but not contractile bundles.

\newpage

\begin{figure}[htbp]
    \centering
    \includegraphics[width=0.5\textwidth]{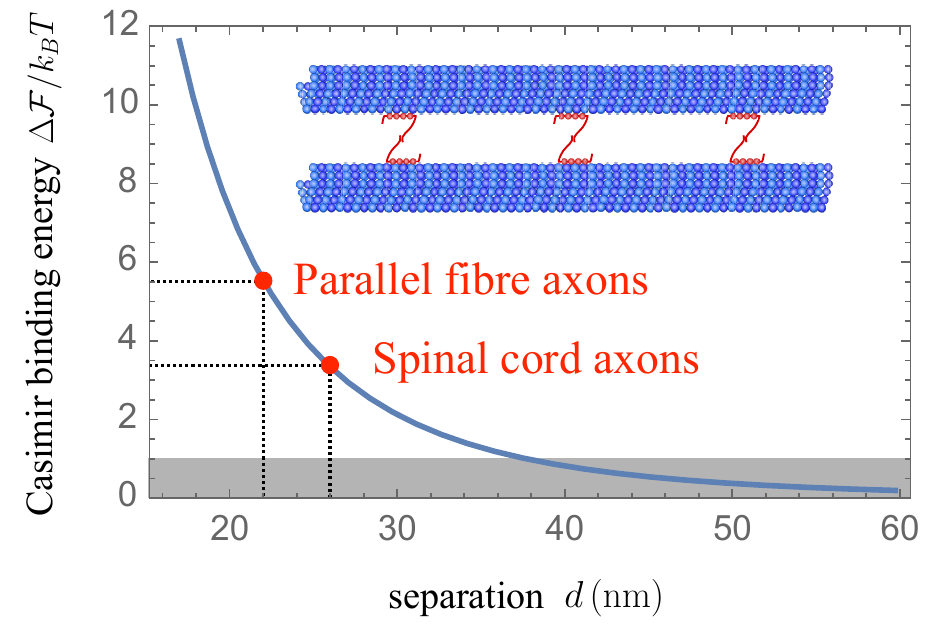}
    \caption{ \textbf{Casimir binding free energy between microtubules.}
    Same as in Fig.~\ref{fig:fig2}, but in the case of two parallel microtubules, each with radius $12\,{\rm nm}$ and length $50\,\mu{\rm m}$.  Inset: schematic of two microtubules (blue) in a bundle. Cross-linkers are shown in red.}
    \label{fig:fig3}
\end{figure}

Another biologically important system where Casimir interactions between filaments are relevant regards microtubule bundles. Microtubules can be viewed as cylinders with a radius of about $12$~nm. They can grow as long as 50~$\mu$m and their persistence length is around 1 mm~\cite{Hawkins10}. They form bundles where the separation between neighboring microtubules is set by microtubule-associated proteins, of which various types exist~\cite{Balabanian18}. Plant cells often possess large arrays of parallel microtubules, whose alignment is maintained over the whole cell, and where separations are similar to the microtubule diameter~\cite{Chan99,Gaillard08}. Microtubule bundles are also present in neurons, where they play important roles, and the separations between adjacent microtubules in Purkinje cell dendrites, parallel fibre axons and white matter spinal cord axons were found to be $64\pm 10$~nm, $22\pm 10$~nm and $26\pm 10$~nm, respectively~\cite{Chen92}.
For $d=22\,{\rm nm}$, we find $\Delta{\cal F}/L \approx 0.112\, k_B T /\mu{\rm m}$, giving $\Delta{\cal F}=5.6\, k_B T$ for $L= 50\,\mu{\rm m}$ and $\Delta{\cal F}=1.7\, k_B T$ for $L= 15\,\mu{\rm m}$. In \cref{fig:fig3}, we show the Casimir binding free energy $\Delta\mathcal{F}$ versus the separation $d$ for microtubules with $L= 50\,\mu{\rm m}$. The Casimir interaction then exceeds the scale of thermal fluctuations for separations $d\lesssim 35$~nm, which includes the physiological separations found in parallel fibre axons and white matter spinal cord axons, as well as in plant cells, but not in Purkinje cell dendrites.

\section*{Discussion}

The long-range part of the Casimir interaction has a universal form between two dielectric objects immersed in salted water. Here, we calculated the Casimir interaction in the case of two long parallel dielectric cylinders. We demonstrated that this interaction takes values substantially larger than the scale of thermal fluctuations, in the important biological cases of actin bundles and microtubule bundles. 

Actin bundles and microtubule bundles are typically held in place by cross-linking proteins in cells. However, in electrolyte solutions containing polycations e.g. Mg$^{2+}$, actin filaments can form bundles \textit{in  vitro} in the absence of cross-linking proteins~\cite{Tang96,Deshpande12}. Microtubule bundles can also self-assemble \textit{in vitro} above a certain concentration of multivalent cations~\cite{Needleman04,Hamon11}. This demonstrates that the tendency of these filaments to self-assemble is quite generic. In this light, cross-linkers could help maintain spacing between filaments~\cite{Mephon16}. Interestingly, beyond cytoskeletal proteins, multiple enzymes form filamentous structures within cells, which then assemble into large-scale self-assembled structures (foci, rods, rings, sometimes called cytoophidia), which are membraneless and reversible. Enzyme filamentation is associated to multiple functions in cells, including determining cell shape or regulating enzyme activity~\cite{Park19}. Enzymes that form filaments and higher-order structures in cells include acetyl CoA carboxylase (ACC), CTP synthetase (CtpS)~\cite{Ingerson10,Barry14}, inositol monophosphate dehydrogenase (IMPDH)~\cite{Juda14,Johnson20}, and many others~\cite{Park19}. Beyond enzymes, it was recently shown that mutating proteins that spontaneously form symmetric homo-oligomers can lead to polymerization in a quite generic manner. Furthermore, these mutant proteins very often form larger structures such as fibers or foci, and some were shown to bundle~\cite{Garcia17,Garcia22}. These findings hint at a general trend of filaments to self-assemble into higher-order structures in cells.

While the universal attractive interaction discussed here should play a part in these various bundles, these systems are complex and involve many other interactions. Electrostatic interactions are screened at the usual separations involved in actin bundles and microtubule bundles, but they matter at shorter separations. For instance, the surface of filamentous actin is overall negative, with a highly heterogeneous charge distribution, leading to subtle collective dynamics of counterions close to actin filaments~\cite{Angelini06}. In addition, the depletion interaction~\cite{Asakura54,Asakura58,Vrij76,Marenduzzo06}, which arises from excluded volume effects on crowding agents, is important in cells due to how crowded the cytoplasm is. Indeed, around 30\% of its volume is estimated to be occupied by macromolecules~\cite{Zimmerman91,Ellis01,Marenduzzo06}, with notable heterogeneities~\cite{Gnutt15,Rivas16}. The range of the depletion interaction is given by the diameter of a typical depletant, which corresponds to macromolecules such as globular proteins in cells, with diameters of order 5~nm~\cite{Marenduzzo06}. Depletion interactions have been studied experimentally in controlled in-vitro systems where the concentration of depletant polymers can be tuned. For instance, an adhesion strength of $7\,k_\mathrm{B}T/\mu$m was measured between sickle hemoglobin fibers in a solution of monomeric hemoglobin~\cite{Jones03}, while the attractive interaction between two actin filaments in a solution of depletant polymers was found to be of order of a few times $10\,k_\mathrm{B}T/\mu$m~\cite{Lau09}, and a similar value was found between two microtubules~\cite{Hilitski15}. Thus, this interaction is strong between parallel filaments in a cell, but it is also very short-ranged, with a range of order 5~nm.

What sets the Casimir interaction we calculated apart from electrostatic and depletion interactions, and to our knowledge, from all other interactions at equilibrium, is its long range, which arises from the lack of screening of transverse electromagnetic fluctuations. It is this long range that makes it quantitatively important e.g. between actin filaments at the physiological separation found in parallel bundles. It also means that this force could play important part in the self-assembly of these structures.

An out-of-equilibrium long-range fluctuation-induced interaction was recently predicted between neutral objects immersed in  electrolytes subject to an external electric field \cite{Golestanian2021}. This force can be of importance at the cell scale, e.g. for ion channels. More generally, Casimir forces present interesting out-of-equilibrium properties~\cite{Dean12, Dean16}. Such effects could be all the more important for cytoskeletal filaments that the cytoskeleton is an active system~\cite{Julicher07,Gladrow16,Gross19}. Here, we showed that an equilibrium long-range universal interaction exists between filaments in cells. 

The Casimir interaction is highly dependent on the geometry of the interacting objects because it arises from the perturbation of electromagnetic fluctuations by the interacting objects. Here, we showed that its magnitude is several times the scale of thermal fluctuations in the geometry of two parallel filaments whose length (micrometer-scale) is much larger than their radius and separation (nanometer-scale). This is stronger than between two spheres~\cite{Schoger2022}, because long cylinders have a stronger confining effects on electromagnetic fluctuations than spheres. Accordingly, within a cell, the Casimir interaction can be strong between long semi-rigid biopolymers such as those considered here, but is weak between globular proteins. Note that similar geometric effects exist for other fluctuation-induced interactions, e.g. in the case of Casimir-like interactions induced by the thermal fluctuations of the shape of the membranes: these interactions are stronger between long parallel rods adsorbed on a membrane~\cite{Golestanian96,Bitbol11} than between circular or point-like inclusions modeling transmembrane proteins~\cite{Goulian93,Fournier97}. Critical Casimir forces can also be important for cylindrical particles immersed in critical binary mixtures~\cite{Trondle10,Labbe14,Labbe17}. In addition to parallel cylinders, another biologically relevant case where the electromagnetic Casimir interaction should matter regards stacks of lipid membranes. Modeling them by parallel dielectric planes immersed in salted water, the Casimir binding free energy between two lipid membranes is $\Delta\mathcal{F}=2.4\times 10^{-2}k_BT\,A/d^2$, where $A$ is the area of the planes and $d$ their separation \cite{MaiaNeto2019}. It should thus oppose the repulsive Helfrich and hydration interactions~\cite{Helfrich78,Cevc1987,leikin1993,Petrache06,Freund12,Wennerstrom14,Lu15}. 

\section*{Methods}

\label{sec:meth}

We consider the Casimir interaction between two parallel dielectric cylinders with the same radius $R$ separated by a distance of closest approach $d$, as illustrated in~\cref{fig:fig1}. The cylinders model filaments in a cell. While the dielectric properties of proteins are complex and constitute an important research topic~\cite{Amin2019,Amin2020,Kim2020}, we find that the force we focus on is universal and does not depend on the detail of these properties, thereby freeing us from choosing a specific model. This holds in the temperature regime appropriate at room or body temperature, and for the distances considered here. We denote by $L$ the (common) length of the cylinders, and assume that $L\gg d$, so that edge effects can be neglected, which allows us to make calculations in the simpler case of infinitely long cylinders. This hypothesis is satisfied by the biological systems of interest. The cylinders are immersed in an aqueous solution (modeling the cytoplasm) with a Debye screening length smaller than the distance of closest approach $d$, which entails that electrostatic interactions between them are screened. For actin filaments or microtubules in the cytoplasm, this hypothesis holds down to a separation of the order of $0.5\,{\rm nm}$~\cite{Spitzer05}. 

We evaluate the Casimir interaction between two filaments immersed in salted water, focusing on the long-range unscreened part associated to the transverse magnetic polarization. This long-range interaction has recently been calculated between planes and between spheres \cite{MaiaNeto2019,Schoger2022}, and these theoretical predictions have been experimentally verified in the case of a small sphere held by optical tweezers in the vicinity of a larger rigidly held sphere \cite{Pires2021}. 
The two-parallel-cylinder geometry studied here is a new situation, and we performed calculations in this case.

Our calculations are based on scattering theory \cite{Lambrecht2006,Rahi2009}, a rigorous approach inspired by quantum optics, where the Casimir force is computed from the radiation pressure of the electromagnetic field fluctuations. This method has been extremely successful for describing a broad variety of configurations~\cite{Reynaud08,Milton2008,Woods2016,Gong2021,Bimonte2022}. The Casimir interaction is written as a sum over Matsubara frequencies \cite{DLP1961,Schwinger1978}. The first term of this sum, corresponding to zero frequency, overtakes all other ones when thermal fluctuations dominate. It is the case in biological systems at room or body temperature, for the distances relevant in bundle formation and cohesion, and therefore we focus on this zero-frequency term, which yields the thermal Casimir interaction energy~\cite{MaiaNeto2019}. This term can be written as an integral over electromagnetic modes, involving the round-trip operator after scattering on the two rods~\cite{MaiaNeto2019,Schoger2021} (see also~\cite{Lambrecht2006,Reynaud08}). The round-trip operator accounts for scattering on each cylindrical rod~\cite{Bohren1998}, as well as for translation from the axis of one rod to the other one and vice-versa.
Because of Debye screening, only transverse modes actually contribute. Moreover, because the dielectric function of salted water diverges in the zero-frequency limit~\cite{Rinne14}, the interaction does not depend on the specific dielectric properties of the filaments. 

We checked that our general result is consistent with the proximity force or Derjaguin approximation~\cite{Israelachvili2011} in the regime $d\ll R$, as well as with the expectation in the long-distance limit $d\gg R$. We stress however that these extreme regimes are not the relevant ones for biological systems, as actin bundles and microtubule bundles are in the intermediate range $d\gtrsim R$ where both short- and long-distance approximations fail by a large factor. Therefore, our full calculations are required to precisely estimate the force in these systems. While the scattering formalism yields an explicit formula for the Casimir force between dielectric cylinders immersed in salted water, its  evaluation involves technical numerical methods. Specifically, we employ a decomposition of electromagnetic fields on the plane wave basis, similar to that used in \cite{Spreng2020} for the interaction between two spheres. The numerical values of the function $\phi$, as well as the code allowing to compute them, are freely available (see Code availability and Data availability).

To shed more light on the fact that transverse modes of the electrostatic field are unscreened, which is the reason for the existance of the long-range part of the Casimir interaction, we also performed molecular dynamics simulations in the two cases of pure water and a KBr electroyte solution using GROMACS 2021 \cite{gromacs}. These simulations allow us to numerically estimate longitudinal and transverse correlations of the field, and to highlight the specific coupling of the salt with longitudinal modes. We further derived the correlations of longitudinal and transverse modes in water and electrolytes, using classical field theory calculations based on nonlocal electrostatics. Details are presented in the Supplementary Material.

\section*{Code availability}
Our code will be freely available in a Zenodo repository upon publication.

\section*{Data availability}
The data in our figures will be freely available in a Zenodo repository upon publication.

\section*{Acknowledgments}
We are grateful to Gert-Ludwig Ingold, Astrid Lambrecht and Tanja Schoger for stimulating discussions. 
A.-F.~B. thanks the European Research Council (ERC) for funding under the European Union’s Horizon 2020 research and innovation programme (grant agreement No.~851173, to A.-F.~B.). 
H.~B. acknowledges funding from the Humboldt Research Fellowship Program for Experienced Researchers.
P.~A.~M.~N. thanks Sorbonne Université for hospitality and
acknowledges funding from the Brazilian agencies Conselho Nacional de Desenvolvimento Cient\'{\i}fico e Tecnol\'ogico (CNPq),
Coordenaç\~ao de Aperfeiçamento de Pessoal de N\'{\i}vel Superior (CAPES),  Instituto Nacional de Ci\^encia e Tecnologia Fluidos Complexos  (INCT-FCx), and the
Research Foundations of the States of Rio de Janeiro (FAPERJ) and S\~ao Paulo (FAPESP). 

\newpage

\renewcommand{\thefigure}{S\arabic{figure}}
\renewcommand{\thetable}{S\arabic{table}}
\renewcommand{\theequation}{S\arabic{equation}}
\setcounter{figure}{0}
\setcounter{equation}{0}
\setcounter{table}{0}

\begin{center}
	\textbf{\huge{Supplementary Material}}
\end{center}
\vspace{0.1cm}

\section{Outline: Transverse electrostatic modes of water are unaffected by salt}
In this supplementary section, we perform classical molecular dynamics (MD) simulations to investigate how salt affects the bulk water-water electrostatic correlations. The work is organized as follows: 
\begin{itemize}
	\item We extract longitudinal and transverse dielectric response functions for water and electrolytes from MD simulation data.
	\item We observe that, contrary to longitudinal ones, the transverse modes of the polarization field are not perturbed by salt.
	\item We develop a classical field theory (FT) to 
	interpret these observations.
	\item We conclude that longitudinal modes of water-water interactions are screened by salt whereas transverse ones remain unaffected.
\end{itemize}

\section{Molecular dynamics computation of the longitudinal and transverse dielectric susceptibilities for pure water and electrolytes}

\subsection{Molecular dynamics simulation methods}
\begin{figure}[h!]
	\includegraphics[scale=0.13]{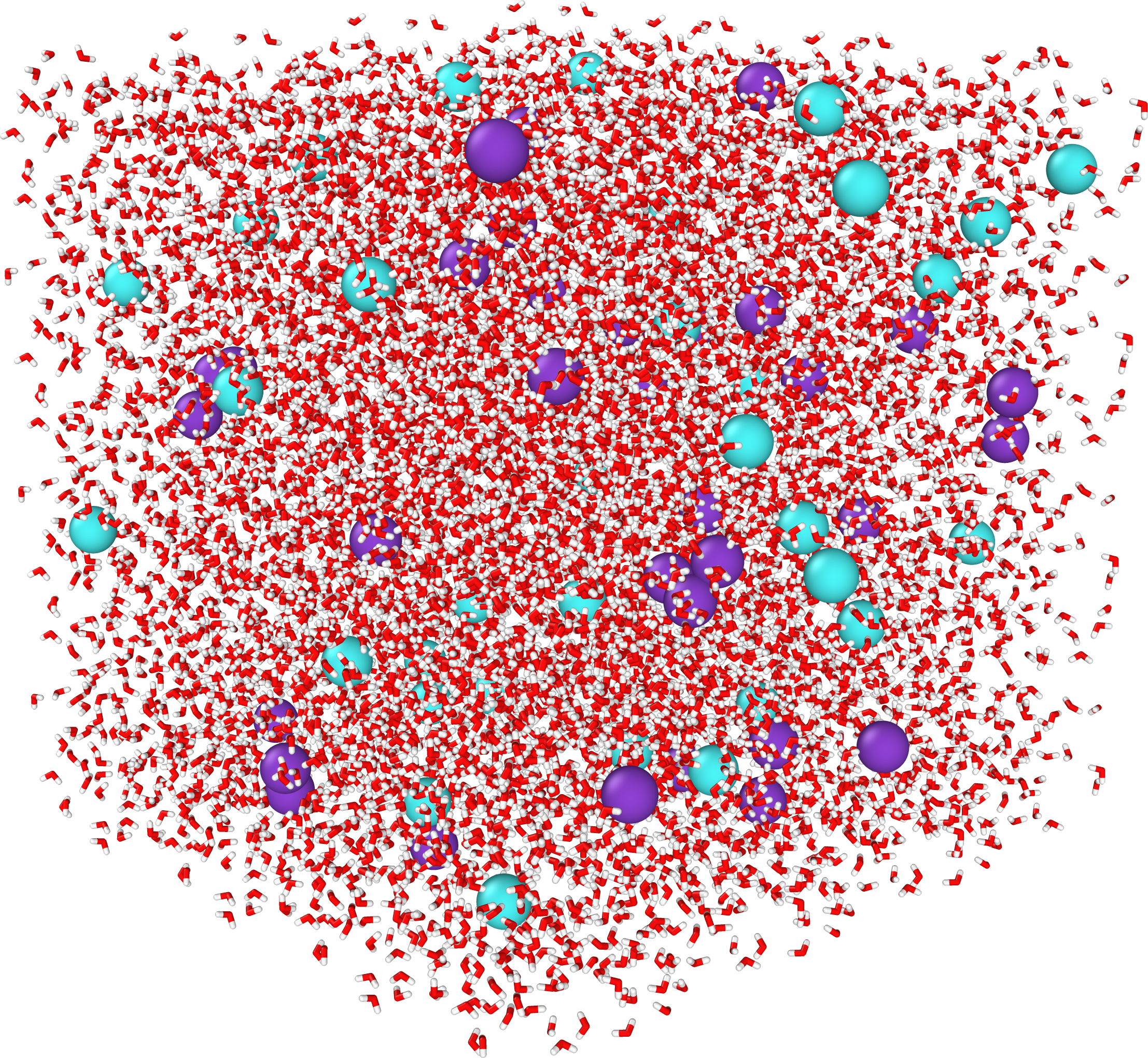}
	\centering
	\caption{\textbf{Snapshot of our MD simulation of an electrolyte solution.} The red and white sticks represent water molecules, the light blue spheres represent potassium K$^+$ ions, and the purple spheres bromide Br$^-$ ions. The box is a cube with side length $L=6.5$~nm. Periodic boundary conditions are used. }
	\label{fig:SI2}
\end{figure}

We consider a cubic water box of side length $L=6.5$~nm composed of $N_w=8967$ water molecules. We simulate both pure water and a 0.2 moles per liter aqueous solution of KBr. The electrolyte solution contains 33 ion pairs in the box. Fig. \ref{fig:SI2} shows the simulation box.
Our focus is on dielectric properties. Thus, rather than simulating a complex solution matching the cytosol composition, we consider a simple solution whose permittivity is well tabulated~\cite{azcatl2014,loche2021} and matches the cytosolic one. Its Debye length, $\lambda_D=0.7$~nm, is in the typical range of biologically relevant solutions.

Simulations are performed using the \texttt{GROMACS 2021} molecular dynamics simulation package~\cite{abraham2015}. The integration time step is set to $\Delta t=2$~fs. Periodic boundary conditions are used in all directions. Long range electrostatics are handled using the smooth particle mesh Ewald (SPME) technique. Lennard-Jones interactions are cut off at a distance $r_{\rm cut}=0.9$~nm.  A potential shift is used at the cut-off distance. All systems are coupled to a heat bath at 300~K using a v-rescale thermostat with a time constant of 0.5~ps. We use \texttt{MDAnanlysis} to treat the trajectories. After creating the simulation box, we perform a first energy minimization. Specifically, we equilibrate the system in the NVT ensemble for 200 ps, and afterwards in the NPT ensemble for another 200 ps using a Berendsen barostat at 1 bar.
Production runs are then performed in the NVT ensemble for 20 ns. 

We performed simulations with the TIP4P/$\epsilon$ water model~\cite{azcatl2014}, a 4 interaction site,  three point-charges and one Lennard Jones reference site model. The Lenard-Jones (LJ) center is placed on the oxygen atom. Charges are placed on the hydrogen atoms and on an additional interaction site, M, carrying the negative charge. The ions (K$^+$ and Br$^-$) were treated according to the force field developed in~\cite{loche2021}.

\subsection{Computation of the susceptibilities}
To compute the dielectric susceptibility $\chi(\vec{q})$ in Fourier space, we use the fluctuation-dissipation theorem, relating $\chi(\vec{q})$ to the fluctuations of the polarization field $\bm{\mathcal{P}}$, as follows,
\begin{equation}
	\label{FDT}
	\langle \bm{\mathcal{P}}(\vec{q}) \bm{\mathcal{P}}(\vec{\vec{-q}}) \rangle=\epsilon_0 k_BT \chi(\vec{q})\,.
\end{equation}
Using the isotropy and the homogeneity of the medium, the susceptibility can be decomposed in a longitudinal part $\chi_\parallel(q)$ and a transverse part $\chi_\perp(q)$ as follows:
\begin{align}
	\chi_{ij}(q)&=\chi_{\parallel}(q)\frac{q_iq_j} {q^2}+\chi_\perp(q)\left(\delta_{ij}-\frac{q_iq_j}{q^2}\right)\quad \textrm{where } (i,j)\in\{x,y,z\}^2\,.
	\label{chiMFmoldyn}
	%\chi_{\parallel}(q)&=&(K_\parallel(q)+1)^{-1}, \quad \chi_\perp(q)=K_\perp^{-1}(q)
\end{align}

\paragraph*{Longitudinal susceptibility.} The local partial charge $\rho$ of water obeys  $\rho(\vec{r})=-\bm{\nabla} \cdot \bm{\mathcal{P}}(\vec{r})$. Using this relation, one can express the longitudinal susceptibility as a function of the charge structure factor $S(q)$:
\begin{equation}
	\chi_\parallel(q)=\frac{S(q)}{q^2 \epsilon_0 k_BT}\,.\label{chiparallel}
\end{equation}
The charge structure factor in Fourier space can be decomposed into an intramolecular and an intermolecular part,
\begin{equation}
	S(q)=S_{\rm intra}(q)+S_\mathrm{inter}(q)\,.
\end{equation}
The intermolecular contribution reads
\begin{equation}
	S_{\rm inter}(q)=\frac{4 n_w z^2 e^2}{q^2}\left[h_{\rm MM}(q)+h_{\rm HH}(q)-2h_{\rm HM}(q)\right]\,,
\end{equation}
where $z$ denotes valency, $e$ the elementary charge, $n_w$ the molecular number density, while $h_{\rm IJ}$ is the Fourier transform of $g_{\rm IJ}(r)-1$, $g_{\rm IJ}(r)$ being the radial distribution function associated with the atom couple IJ. 
Next, the intramolecular contribution can be written as
\begin{equation}
	S_{\rm intra}(q)=\frac{4 n_w z^2 e^2}{q^2}\left(\frac{\sin(q d_{\rm HH})}{q d_{\rm HH}}-4\frac{\sin(q d_{\rm HM})}{q d_{\rm HM}}+3\right)
\end{equation}
where $ d_{\rm IJ}$ is the intramolecular distance between atoms I and J. 
At low $q$, the accuracy of this expression of the structure factor decreases, because the function $h_{\rm IJ}(r)$ is obtained at a finite range, imposed by the box size.  To solve this problem, we proceed as follows. 
For $q< 2.5$ \AA$^{-1}$, we take into account the periodicity of the system by calculating the charge structure factor for discretized values of the wave vector norm $q$, namely $q=2\pi/L\sqrt{n_x^2+n_y^2+n_z^2}$, where $(n_x,n_y,n_z)$ are non-negative integers. We then compute directly the charge structure factor from the charge distribution $	\tilde{\rho}(q)$ in Fourier space, which reads 
\begin{eqnarray}
	\label{rhoq}
	%	\rho(r)&=&\Sigma_{i=1}^N q_H\left(-2\delta(r-r_{\rm O,i})+\delta(r-r_{\rm H1,i})+\delta(r-r_{\rm H2,i})\right)\nonumber\\
	\tilde{\rho}(q)&=&\sum_{i=1}^{N_w} e\,z\,  e^{i \vec{q} \cdot \vec{r}}\left(-2 e^{-i\vec{q}.\vec{r}_{{\rm M}_{i}}}+e^{-i\vec{q}.\vec{r}_{{\rm H}_{1,i}}}+e^{-i\vec{q}.\vec{r}_{{\rm H}_{2,i}}} \right)
\end{eqnarray}
where H$_{1,i}$ and H$_{2,i}$ stand for the two hydrogen atoms of molecule $i$. The charge structure factor is then obtained as
\begin{eqnarray}
	S(q) &=&\frac{1}{V}\left\langle \tilde{\rho}(q) \tilde{\rho}(-q) \right\rangle\\&=&\frac{2q_H^2}{V}\sum_i\sum_{j\leq i}\Big[4 \cos(\vec{q} \cdot \vec{d}_{\rm M_iM_j})-2\cos(\vec{q} \cdot \vec{d}_{\rm M_iH_{1,j}})-2\cos(\vec{q} \cdot \vec{d}_{\rm H_{1,i}M_j})-2\cos(\vec{q} \cdot \vec{d}_{\rm M_iH_{2,j}})\nonumber\\&-&2\cos(\vec{q} \cdot \vec{d}_{\rm H_{2,i}M_j})+\cos(\vec{q} \cdot \vec{d}_{\rm H_{1,i}H_{1,j}})+\cos(\vec{q} \cdot \vec{d}_{\rm H_{2,i}H_{2,j}})+\cos(\vec{q} \cdot \vec{d}_{\rm H_{1,i}H_{2,j}})\nonumber\\&+&\cos(\vec{q} \cdot \vec{d}_{\rm H_{2,i}H_{1,j}}))\Big]\,,
\end{eqnarray}
where $q_H$ is the charge of the hydrogen atom, $\vec{q}$ is the wave vector, while $\vec{d}_{\rm{A}_i\rm{A}_j}$ stands for $\vec{r}_{\rm{A}_i}-\vec{r}_{\rm{A}_j}$, and $V$ denotes the volume of the simulation box. The longitudinal susceptibility can then be obtained from Eq.~(\ref{chiparallel}).\par
\begin{figure}[h!]
	\includegraphics[scale=1]{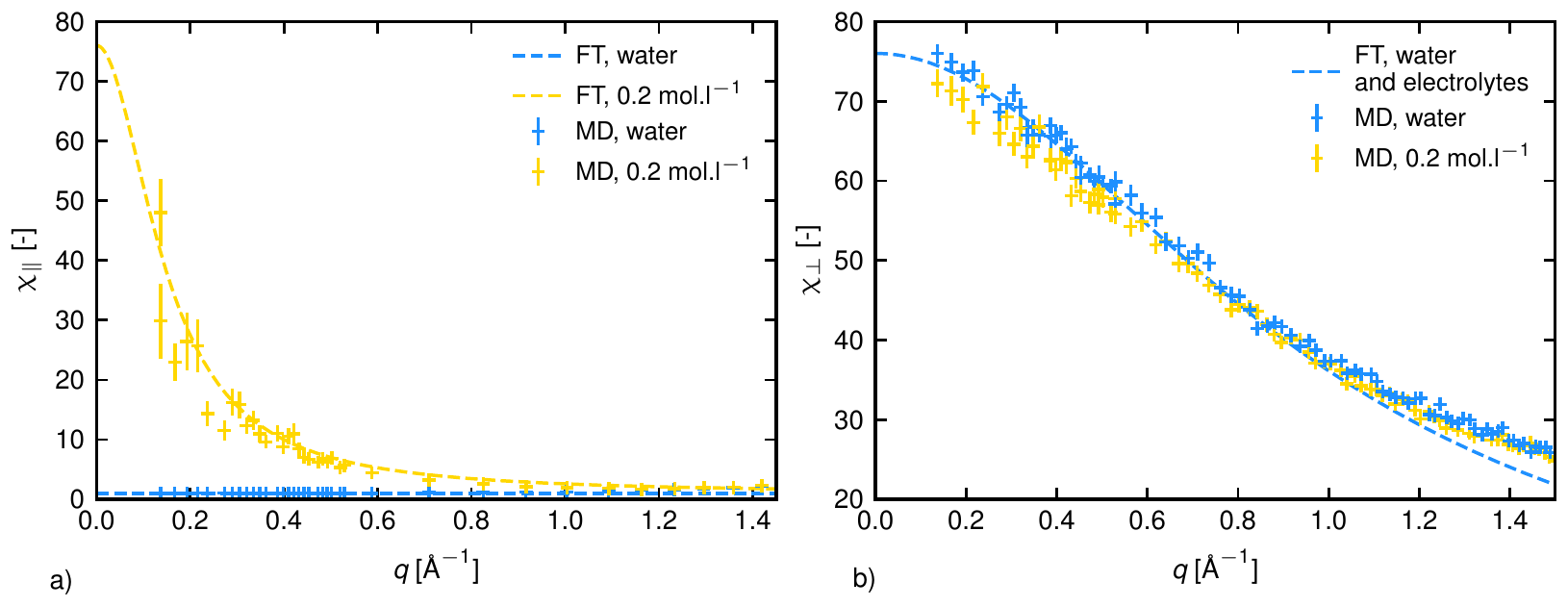}
	\centering
	\caption{\textbf{Susceptibilities for water and electrolytes.} a) Longitudinal susceptibility $\chi_\parallel(q)$. b) Transverse susceptibility $\chi_{\perp}(q)$. In both panels, susceptibility is shown as a function of the wave vector norm $q$ (\AA$^{-1}$) for pure water (blue curve) and for a 0.2 mole per liter KBr electrolyte solution (yellow curve). Markers correspond to results obtained from MD simulations (see text for details). Dashed lines correspond to FT equations given respectively by Eqs. (\ref{chiMFmoldyn}) and (\ref{chisalt}) for pure water and for the electrolyte, for parameter values $K=1/76$, $\kappa_c=0.0145$~\AA$^{2}$. The Debye length associated to the electolyte solution considered here is $\lambda_D=6.8$~\AA.}
	\label{fig:SI}
\end{figure}

%\begin{eqnarray}
%\langle  \rho(q)\rho(-q) \rangle&=&q_H^2\Sigma_{i,j, j\neq i}\Big(4 \cos(q \cdot d_{\rm OiOj})-4\cos(q \cdot d_{\rm OiH1j}))\nonumber\\&-&4\cos(q \cdot d_{\rm OiH2j})+\cos(q \cdot d_{\rm H1iH1j})+\cos(q \cdot d_{\rm H2iH2j})\nonumber\\&+&2\cos(q \cdot d_{\rm H1iH2j})+\cos(q \cdot d_{\rm H2iH1j}))\Big)
%\end{eqnarray}
We report in Fig. \ref{fig:SI} a) the longitudinal susceptibility for pure water (blue markers) and for the electrolyte (yellow markers) for $q\leq1.5~$\AA$^{-1}$ as we focus on long-range interactions. We observe that the longitudinal susceptibility of the electrolyte significantly differs from the pure water one at low $q$.

\paragraph*{Transverse susceptibility.} The transverse susceptibility is computed following~\cite{kornyshevtrans}.
The polarization of the medium in Fourier space, namely
\begin{equation}
	{\bf P}({\bf q})=\Sigma_j {\bf p}_j({\bf q})e^{-i {\bf q}\cdot  {\bf r}_j}\,,
\end{equation}
can be written as a sum over the molecular polarization ${\bf p}_j({\bf q})$ of molecule $j$, which reads
\begin{equation}
	{\bf p}_j({\bf q})=	\frac{1}{\sqrt{V}}\sum_\alpha \frac{e\, z_\alpha\, {\bf \delta r}_{\alpha j}}{i\, {\bf q}\cdot {\bf \delta r}_{\alpha j } }\left(1-e^{-i\,{\bf q}\cdot {\bf \delta r}_{\alpha j }}\right)\,,
\end{equation}
where ${\bf \delta r}_{\alpha j } $ denotes the distance between the charge $\alpha$ and the center of mass of the molecule. 
We then take the transverse part of the polarization ${\bf P}_\perp({\bf q})={\bf q}\times {\bf P}(\bf q)/q$, and compute the transverse susceptibility as
\begin{equation}
	\label{chiperpMD}
	\chi_\perp(q)=\frac{\langle{\bf P}_\perp({\bf q})\cdot{\bf P}_\perp({\bf -q})\rangle}{k_BT\epsilon_0}.
\end{equation}
Note that we replace $\left(1-e^{-i{\bf q}\cdot {\bf \delta r}_{\alpha j }}\right)/(i\,{\bf q}\cdot {\bf \delta r}_{\alpha j }) $ by 1 if $ {\bf q}\cdot {\bf \delta r}_{\alpha j } <10^{-5}$ to prevent numerical errors.\par 
We report in Fig.~\ref{fig:SI} b) the transverse susceptibility for water (blue markers) and for electrolyte (yellow markers) for $q\leq1.5\,$\AA$^{-1}$. We observe that the transverse susceptibility is not affected by the salt. 

\subsection{Statistical treatment}
For the longitudinal and transverse susceptibilities, the error bars were derived following the reblocking method \cite{flyvbjerg1989}.
For the bulk permittivity, we cut the trajectory in 5 statistically independent blocks, compute the bulk permittivity of each block,  estimate the sample variance $\sigma^2$ and define the error bar as $\sqrt{\sigma^2/5}$.

\section{Classical field theory interpretation}

\subsection{Water as a nonlocal dielectric medium}
To better understand why longitudinal and transverse fluctuations are  differently affected by salt, we develop a classical field theory model for water and electrolytes.
We describe water as a continuous nonlocal and linear dielectric medium~\cite{maggs2006}. The electrostatic energy $\mathcal{U}_{\rm el}$ of the medium is written as a functional of the polarization field ${\bm{\mathcal P}}$ as follows:
\begin{equation}
	\label{HP}
	\mathcal{U}_{\rm el}[\bm{\mathcal{P}}]=  \frac{1}{2}\int d{\vec r} d{\vec r} '\frac{\bm{\nabla}_{\vec{r}} \cdot \bm{\mathcal{P}}(\vec{r})\bm{\nabla}_{\vec{r}'}   \cdot \bm{\mathcal{P}}(\vec{r}')}{4 \pi \epsilon_0|\vec{r}-\vec{r}'|}+\frac{1}{2 \epsilon_0}\int d\vec{r} \bm{\mathcal{P}}(\vec{r})\cdot K(\vec{r}-\vec{r}')\cdot \bm{\mathcal{P}}(\vec{r}')\,.
\end{equation}
The first term is the Coulomb energy, where we expressed the partial charge $\rho_w$ in the fluid as $\rho_w(\vec{r})=-\nabla_r \cdot \bm{\mathcal{P}}(\vec{r})$. The second term corresponds to a conformation energy of the medium whose kernel $K$ depends on $\vec{r}-\vec{r}'$ assuming isotropy and homogeneity. It can be expanded following a Landau-Ginzburg approach to encode the correlations of the fluid at the nanoscale. 
The dielectric susceptibility $\chi(\vec{r}-\vec{r}')$ of the system is defined as 
\begin{eqnarray}
	\mathcal{U}_{\rm el}[\bm{\mathcal{P}}]&=&\frac{1}{2\epsilon_0}\int d\vec{r} d\vec{r}'\bm{\mathcal{P}}(\vec{r}) \cdot \chi^{ -1}(\vec{r}-\vec{r}')\cdot \bm{\mathcal{P}}(\vec{r}'),\nonumber\\
	\label{chiparaperp}
	&=&\frac{1}{2\epsilon_0}\int d\vec{r} d\vec{r}'\left(\bm{\mathcal{P}}_\parallel(\vec{r}) \cdot\chi_\parallel^{ -1}(\vec{r}-\vec{r}')\cdot \bm{\mathcal{P}}_\parallel(\vec{r}')+\bm{\mathcal{P}}_\perp(\vec{r}) \cdot\chi_\perp^{ -1}(\vec{r}-\vec{r}')\cdot \bm{\mathcal{P}}_\perp(\vec{r}')\right),
\end{eqnarray}
where we have split the polarization field into a longitudinal part $\bm{\mathcal{P}}_\parallel$ and a transverse part $\bm{\mathcal{P}}_\perp$, which respectively satisfy 
\begin{equation}
	\bm{\nabla}_{\vec{r}}\times  \bm{\mathcal{P}}_\parallel(\vec{r})=0\,,\quad \quad  \bm{\nabla}_{\vec{r}} \cdot  \bm{\mathcal{P}}_\perp(\vec{r})=0\,. 
\end{equation}
The susceptibility can be written in Fourier space as $\chi_{ij}(q)=\chi_{\parallel}(q)\frac{q_iq_j} {q^2}+\chi_\perp(q)\left(\delta_{ij}-\frac{q_iq_j}{q^2}\right).$
We adopt a kernel that captures the constant behavior of the longitudinal response of water and the Lorentzian-like decay of the transverse response (as observed in Fig. \ref{fig:SI}):
\begin{equation}
	\label{Kmodel}
	\bm{\mathcal{P}}(\vec{r})\cdot K(\vec{r}-\vec{r}') \cdot\bm{\mathcal{P}}(\vec{r}')=\left[K \bm{\mathcal{P}}(\vec{r})^2+\kappa_c(\bm{\nabla} \times  \bm{\mathcal{P}}(\vec{r}))^2\right]\delta(\vec{r}-\vec{r}')\,,
\end{equation}
where $K$ is a parameter defining the bulk ($q=0$) properties of the medium, and $\kappa_c$ a Landau-Ginzburg parameter encoding the transverse correlation length of the fluid. 
Using Eq. (\ref{Kmodel}) and inversing Eq. (\ref{chiparaperp}), we find that the longitudinal and transverse susceptibilities are equal to 
\begin{equation}
	\label{chiMF}
	\chi_\parallel(q)=\frac{1}{1+K}, \quad  \chi_\perp(q)=\frac{1}{K+\kappa_cq^2}.
\end{equation}
\subsection{Parametrization of the model using MD simulations}
Fig. \ref{fig:SI} compares results for pure water obtained with MD simulations and with the model presented in Eq. (\ref{chiMF}). On panel a), the longitudinal susceptibility $\chi_{\parallel}$ is plotted as a function of $q$. The longitudinal susceptibility is found to be constant and equal to the bulk susceptibility in MD simulations (blue markers). Furthermore, it is well described by Eq. (\ref{chiMF}) (dashed blue line). Note that the permittivity of the medium  obeys 
\begin{equation}\label{epsilonw}
	\epsilon_w=\left(1-\chi_\parallel(0)\right)^{-1}=1+1/K\,. 
\end{equation}
Panel b)  presents the transverse susceptibility $\chi_{\perp}(q)$. The two parameters of the FT model, namely $K$ and $\kappa_c$, were adjusted to reproduce MD simulations (dashed blue line). The model then predicts well the behavior observed at low $q$ in MD simulations (blue markers).  
\subsection{Response function of electrolytes in the field theory framework}
The partition function for $N_+$ monovalent cations and $N_-$ monovalent anions of respective charges $e$ and $-e$ solvated in this medium can be written as
\begin{eqnarray}
	\mathcal{Z}&=&\frac{1}{N_+!}\frac{1}{N_-!}\left[\prod_{i=1}^{N+}\int d\vec{r}_i \right]\left[\prod_{j=1}^{N-}\int d\vec{r}_j \right]\int \mathcal{D}[\bm{\mathcal{P}}]\,\exp\left[-\frac{\beta}{2\epsilon_0}\int d\vec{r} d\vec{r}' 	\bm{\mathcal{P}}(r)\cdot K(\vec{r}-\vec{r}') \cdot \bm{\mathcal{P}}(r')\right]\nonumber\\
	&\times& \exp\left[-\frac{\beta}{2}\int d\vec{r}\int d\vec{r}'\left[\rho_i(\vec{r}) - \nabla_\vec{r}\cdot \bm{ \mathcal{P}}(\vec{r})\right]v(\vec{r}-\vec{r}')\left[\rho_i(\vec{r}') - \bm{\nabla}_{\vec{r}'}\cdot\bm{ \mathcal{P}}(\vec{r}')\right]\right]\,,
\end{eqnarray}
where $v(\vec{r}-\vec{r}')=1/(4\pi\epsilon_0|\vec{r}-\vec{r}'|)$ denotes the Coulomb potential, while 
\begin{equation}
	\label{rho}
	\rho_i(\vec{r})=\sum_{i=1}^{N+} e\, \delta(\vec{r}-\vec{r}_i)-\sum_{j=1}^{N-} e\, \delta(\vec{r}-\vec{r}_j)
\end{equation}
denotes the ionic charge density.
Introducing an auxilary field $\Phi$ and performing a Hubbard-Stratonovich transform to get rid of the long-range Coulomb potential~\cite{levy2020} yields 
\begin{eqnarray}
	\mathcal{Z}&=&\int\mathcal{D}[\Phi]\frac{1}{N_+!}\left(\int d\vec{r} e^{-i\beta e \Phi(\vec{r})}\right)^{N_+}\frac{1}{N_-!}\left(\int d\vec{r} e^{i\beta e \Phi(\vec{r})}\right)^{N_-}e^{-\frac{\beta}{2}\int d\vec{r} \epsilon_0(\bm{\nabla} \Phi)^2}\nonumber \\
	&\times& \int\mathcal{D}[\bm{\mathcal{P}}]e^{-\frac{\beta}{2}\int d\vec{r} d\vec{r}' 	\bm{\mathcal{P}}\cdot(\vec{r})K(\vec{r}-\vec{r}')\cdot \bm{\mathcal{P}}(\vec{r}')} e^{-i\int d\vec{r} \Phi(\vec{r})(\bm {\nabla} \cdot \bm{ \mathcal{P}}(\vec{r})-\rho_i(\vec{r}))}\,,
\end{eqnarray}
where we have dropped a prefactor.
The partition function is brought into a more manageable form by shifting to the grand-canonical ensemble:
\begin{eqnarray}
	\Xi&=&\sum_{N_+=0}^{\infty}\frac{e^{-\beta \mu_+ N_+}}{N_+!}\left(\int d\vec{r}e^{-\beta e i\phi(\vec{r})}\right)^{N_+}\sum_{N_-=0}^{\infty}\frac{e^{-\beta \mu_- N_-}}{N_-!}\left(\int d\vec{r}e^{\beta e i\phi(\vec{r})}\right)^{N_-}\nonumber\\
	&\times&\int \mathcal{D}[\bm{\mathcal{P}}]e^{-\frac{\beta}{2\epsilon_0} \int d\vec{r} d\vec{r}'\bm{ \mathcal{P}}(\vec{r})\cdot K(\vec{r}-\vec{r}')\cdot \bm{ \mathcal{P}}(\vec{r'})}
	\int \mathcal{D}[\Phi] e^{-\frac{\beta}{2}\int d\vec{r} \epsilon_0\left(\nabla \Phi(\vec{r})\right)^2
		-\int d\vec{r} i\Phi(\vec{r})\left(\bm{\nabla} \cdot {\bm{ \mathcal{P}}}(\vec{r})-\rho_i(\vec{r})\right)},
\end{eqnarray}
where $\mu_+$ and $\mu_-$ denote the chemical potentials of the ions.  Next, we introduce the field $\Psi=i\Phi$, which can be identified as the electrostatic potential~\cite{levy2012}, and we perform the sums over $N_\pm$, yielding
\begin{equation}
	\label{GrandPartFunc}
	\Xi=\int \mathcal{D}[\bm{\mathcal{P}}]\,\mathcal{D}[\Psi]e^{ - \beta F_u[\Psi, \bm{{\mathcal{P}}}]} \,,
\end{equation}
where we have defined the action 
\begin{eqnarray}
	\label{action}
	F_u[\Psi,{\bm {\mathcal P}}]&=& \frac{1}{2\epsilon_0}\int d\vec{r}d\vec{r'}\bm{\mathcal{P}}(\vec{r})\cdot K(\vec{r}-\vec{r}')\cdot \bm{\mathcal{P}}(\vec{r}')-\frac{2n}{\beta}\int d\vec{r} \cosh(\beta e \Psi )\nonumber\\&-&\frac{1}{2}\int d\vec{r} \left[\epsilon_0 (\bm{\nabla}_{\vec{r}} \Psi(\vec{r})^2-2 \Psi(\vec{r}) \bm{\nabla}_{\vec{r}} \cdot \bm{\mathcal{P}}(\vec{r})\right]\,,
\end{eqnarray}
with $n$  the ionic density defined as $n=c\, \mathcal{N}_a$, $c$ being the electrolyte concentration and $\mathcal{N}_a$ the Avogadro number.  The mean fields ($\psi $, $\bm{P}$) are solutions of the equations
\begin{equation}
	\left.\frac{\delta F_u}{ \delta \Psi}\right|_{ \bm{\mathcal{P}}}(\psi, \bm{P})=0, \quad \left.\frac{\delta F_u}{ \delta \mathcal{P}_i}\right|_{\Psi} (\psi , \bm{P})=0\,.
\end{equation}
$\psi$ and $\bm{P}$ both vanish. 
The inverse susceptibility of the medium is given by:
\begin{eqnarray}
	\label{DefSusc}
	\left(\begin{array}{cc} \epsilon_0\chi^G & \chi^{\rm G}_{P,\psi} \\ \chi^{\rm G}_{\psi, P} & \frac{\chi^{\rm G}_{\psi,\psi}}{\epsilon_0}
	\end{array}\right)(\vec{r}_1,\vec{r}_2)= \left(\begin{array}{cc}
		\frac{\delta^2 F_u}{\delta {\bm{\mathcal{P}}}_i(\vec{r}_1) \delta{\bm{\mathcal{P}}}_{j}(\vec{r}_2)} (\psi, \bm{P})   & \frac{\delta^2 F_u}{\delta {\bm{\mathcal{P}}}_{i}(\vec{r}_1) \delta \Psi(\vec{r}_2)} (\psi, \bm{P})\\
		\frac{\delta^2 F_u}{\delta \Psi(\vec{r}_1) \delta {\bm{\mathcal{P}}}_{i}(\vec{r}_2)} (\psi, \bm{P}) & \frac{\delta^2 F_u}{\delta \Psi(\vec{r}_1) \delta \Psi(\vec{r}_2)}(\psi, \bm{P})
	\end{array}\right) ^{-1}.
\end{eqnarray}
Performing the functional derivative of $F_u$ in Fourier space and inverting the matrix, we obtain $\chi_{\psi,P}=i(q_x,q_y,q_z)$ and $\chi_{P,\psi}^T=\chi_{\psi,P}$
and
\begin{eqnarray}
	\chi_{\psi,\psi}(q)&=&-\frac{K}{\frac{1}{\epsilon_e-1}\left(\frac{\epsilon_w}{\lambda_D^2}+q^2\right)+q^2} \\
	\label{chisalt}
	\chi_{ij}(q)&=&\chi_\parallel(q)\frac{q_iq_j}{q^2}+\chi_\perp\left(\delta_{ij}-\frac{q_iq_j}{q^2}\right) , \quad
	\chi_\parallel(q)=\frac{\epsilon_w-1}{\epsilon_w}\frac{\frac{\epsilon_w}{\lambda_D^2}+q^2}{\frac{1}{\lambda_D^2}+q^2}, \quad \chi_\perp(q)=\frac{1}{K+\kappa_t q^2}
\end{eqnarray}
where $\lambda_D=\sqrt{\epsilon_0\epsilon_w/2\beta n e^2}$ is the Debye length, with $\epsilon_w$ in Eq.~(\ref{epsilonw}). These expressions show that the longitudinal susceptibility of the medium is now a function of the salt concentration via the Debye length. Conversely, the transverse susceptibility is not affected by the presence of the salt. 
\subsection{Comparison with MD simulations and conclusion}
MD (markers) and FT (lines) results are represented in Fig. \ref{fig:SI}. On Panel a), the two descriptions consistently show that the longitudinal permittivity of the electrolyte solution strongly deviates from the pure water one at low $q$. We observe a Lorentzian decay of this longitudinal susceptibility  with a typical wave-mode given by the Debye wave-mode $q_D\approx1/\lambda_D=0.15\,$ \AA$^{-1}$. By contrast, on panel b), both FT and MD show that the transverse susceptibility of this saline solution has the same behavior as the pure water one. In both cases, our MD simulation results are in good agreement with the predictions of our FT model. \par 
Overall, these results confirm that the transverse electrostatic modes of water are not screened by salt, contrary to the longitudinal ones that couple with the Debye wave mode.

\newpage

\bibliographystyle{unsrt}

\end{document}